\documentclass[floatfix,showpacs,showkeys,twocolumn,twoside,iop]{revtex4}
\usepackage{amssymb}
\usepackage{amsmath}
\usepackage{textcomp}
\usepackage{rcs}
\RCS $Date$
\usepackage{comment}
\usepackage[pdftex]{graphicx}

\newcommand{\be}{\begin{eqnarray}} 
\newcommand{\ee}{\end{eqnarray}} 
\newcommand{\bml}{\begin{multline}}
\newcommand{\eml}{\end{multline}}

\begin{document}

\title{Spectral analysis on a phonon spectral function of a solid-state plasma in a doped semiconductor
}
\author{Kyung-Soo Yi}
\email[Corresponding author. e-mail: ]{ksyi@pusan.ac.kr}
\author{Hye-Jung Kim}
\altaffiliation{Present address: Department of Physics and EHSRC, University of Ulsan, Ulsan 680-749, Republic of Korea}
\affiliation{Department of Physics,
Pusan National University, Busan 609-735, Republic of Korea}
\date{\RCSDate}

\begin{abstract}
We report an analysis on a phonon spectral function of a solid-state plasma formed in a doped semiconductor. Real and imaginary parts of phonon propagators are evaluated including carrier screening effects within a random phase approximation, and finite-temperature spectral behavior of the phonon spectral function is examined in terms of plasmon--phonon coupled modes and quasiparticle excitation mode of the plasma. 
The results are applied to the case of conduction electrons in a wurtzite GaN considering carrier-phonon coupling channel via polar optical phonons.
We show that the dispersion relations of the plasmon-LO phonon coupled (`upper' and `lower') modes and the character of the additional modes via single quasiparticle excitations are heavily associated with the nonlocal and dynamic behavior of the energy shift and collisional broadening of the dressed phonon propagator of the plasma.

\end{abstract}
\maketitle

\vspace{-0.5cm}
\section{Introduction}

It is well known that polar phonons and plasmons are strongly coupled  in a solid-state plasma through their macroscopic electric fields \cite{Abstreiter1984}.
The mode coupling problems between the longitudinal optic (LO) phonons and plasmons have been investigated extensively both theoretically and experimentally including experimental confirmations in Raman scattering measurements \cite{Pinczuk1977,Abstreiter1979,Romanek}.  
If the frequency of plasma is comparable to that of the LO phonon, the interaction of these excitations is maximized leading to the formation of coupled plasmon—-phonon modes \cite{Quinn-Yi}.

Phonon spectral function of a solid-state plasma reveals the behaviors of coupled plasmon-phonon modes through carrier-phonon interaction in addition to the individual properties of phonons and carriers.
Jain and co-workers reported a low energy quasiparticle excitation-like (QPE-like) mode and proposed its role on the hot electron energy loss rate in a single component electron plasma at low temperature \cite{Jain1988}.
However, no detail has been known about the nature of the low energy QPE-like mode and the spectral behavior of the mode.

In this paper, we present an analysis of LO phonon spectral function of a solid-state plasma illustrating the spectral behaviors of both modes of plasmon-phonon coupled `upper' and `lower' branches \cite{Abstreiter1984} and the QPE-like mode in detail applied to the case of conduction electrons in a doped GaN. 
In GaN, much enhanced carrier-polar phonon couping is expected, compared to GaAs, due to its higher ionicity.
We find that the behavior of the QPE-like mode is heavily associated with the frequency and wavenumber dependent energy shift and collisional broadening of the dressed phonon propagator of the plasma.

\section{Formulation}
The phonon spectral function $\mathcal{A}$ is defined by 
$\mathcal{A}(\vec q,\omega)= -\frac{1}{\pi}\mathcal{I}m D (\vec q,\omega)$,
where $D (\vec q,\omega)$ is the retarded phonon propagator \cite{Bruus,Mahan}.
In the presence of electron-electron interaction in many carriers system, dielectric screening renormalizes the electron--phonon coupling resulting in the dressed phonon propagator $D$ written as \cite{Bruus,Mahan72,Mahan,Jain1988}
\begin{align}
D(\vec q,\omega)= & \frac{2\omega_{\vec q}}{\omega^2-  \omega_{\vec q}^2 - 2 \omega_{\vec q}\mid M_{\vec q}\mid^2 \Pi(\vec q,\omega)/\hbar}.
\label{dressed phonon propagator}
\end{align}
Here $\omega_{\vec q}$ and $M_{\vec q} ( =M_{-\vec q}^{*}$) are the bare (undoped crystal) phonon frequency and the matrix element of specific electron--phonon coupling, respectively \cite{ksyi2007,Mq}, and $\Pi(\vec q,\omega)$ denotes the full retarded polarization function of the plasma. 
The poles of $D(\vec q,\omega)$ determine the renormalized phonon dispersion relations.
In Eq.(\ref{dressed phonon propagator}), $\mid M_{\vec q}\mid^2 \Pi(\vec q,\omega)/\hbar$ in the denominator represents the (complex numbered) phonon self-energy correction [=$\Delta(\vec q,\omega)-i\Gamma(\vec q,\omega)$] via plasma polarization function. 
This collisional broadening modifies the spectral behavior of the phonon spectral function of the material in the $\omega-q$ space.
For the bare phonons in undoped material, we consider a simplified Einstein-type non-dispersive model for polar optical phonons of $\omega_{\vec q}=\omega_{\rm LO} (=92~ \rm meV$ in undoped GaN).

%

\begin{figure*}[th]
\includegraphics[width=14cm]{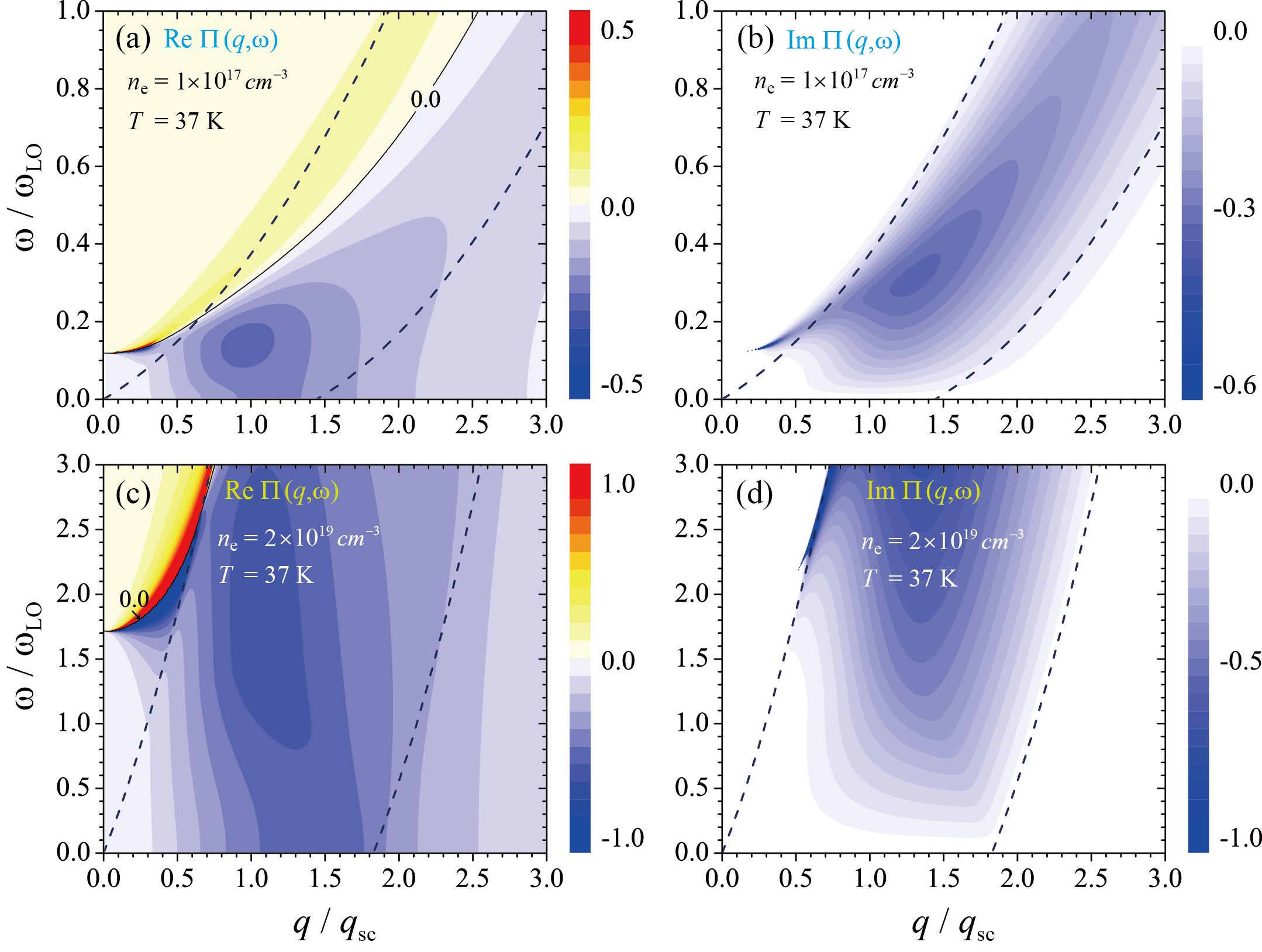}
\caption{Real and imaginary parts of dressed polarization functions, $\Pi(q,\omega)$ at carrier temperature $T$=37 K for a conduction electron plasma of carrier concentration of
 $10^{17} \rm cm^{-3}$ [(a) and (b)] and $2 \times 10^{19} \rm cm^{-3}$ [(c) and (d)].
The wavenumber and frequency are measured in units of the Thomas-Fermi screening wavenumber $q_{\rm sc}$ and the bare LO phonon frequency $\omega_{\rm LO}$.  
Pair of dashed lines denotes the region of allowed quasiparticle excitations in a wurtzite GaN. The zero of ${R}e \Pi (\vec q,\omega)$ is indicated by a thin solid line in panels (a) and (c). }
\label{Pi_37K}
\end{figure*}
\begin{figure*}
\includegraphics[width=14cm]{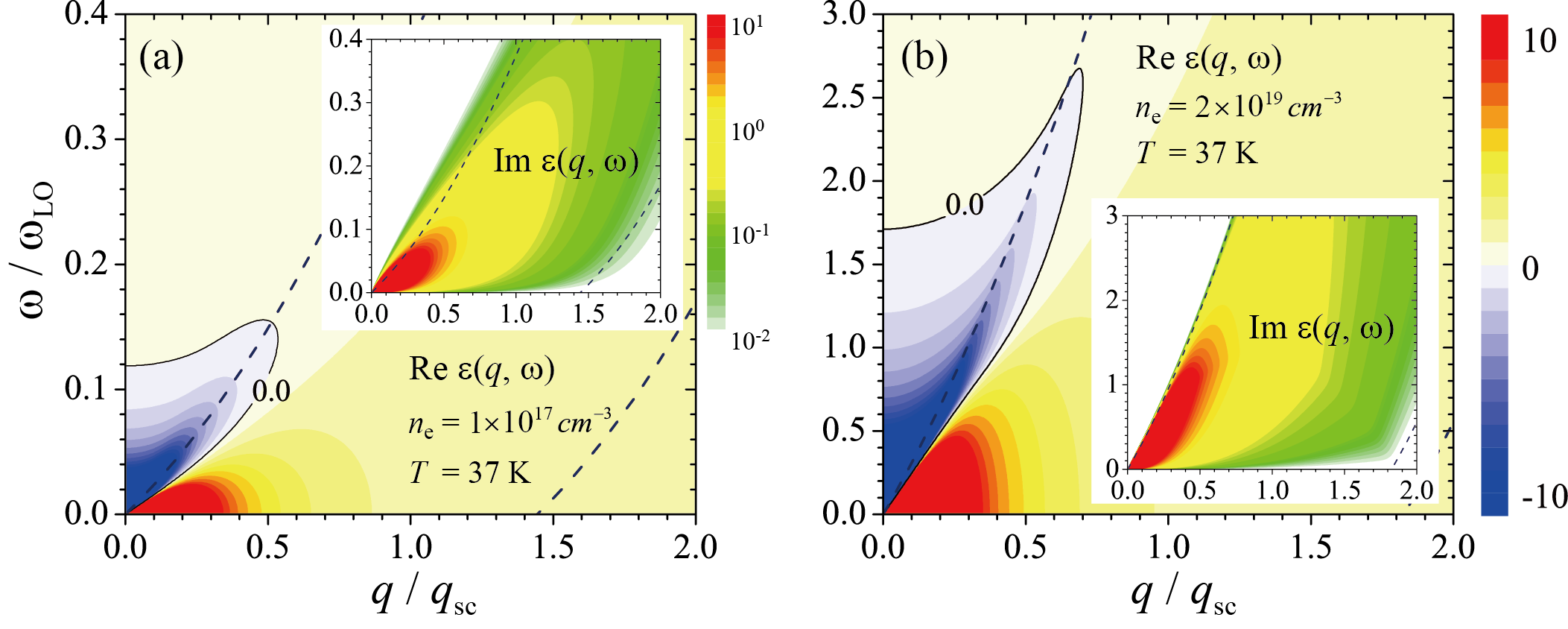}
\caption{Real and imaginary parts of the dielectric function, $\varepsilon(q,\omega)$ at carrier temperature $T$=37 K for a conduction electron plasma of carrier concentration of 
(a) $10^{17} \rm cm^{-3}$ and (b) $2 \times 10^{19} \rm cm^{-3}$.
The zero of ${R}e ~\varepsilon (\vec q,\omega)$ is indicated by thin solid lines in panel (a) and (b). }
\label{epsilon_37K}
\end{figure*}

%

The full retarded polarization $\Pi(\vec q,\omega)$ in Eq.(\ref{dressed phonon propagator}) satisfies Dyson equation given, in terms of irreducible counterpart $\tilde{\Pi}(\vec q,\omega)$, by \cite{Bruus}  
\be
\Pi(\vec q,\omega)=\frac{\tilde{\Pi}(\vec q,\omega)}{1-v_q \tilde{\Pi}(\vec q,\omega)}. 
\label{Pi dyson}
\ee
Here $v_q=4\pi e^2/q^2$, the bare Coulomb interaction. 
The dressed interactions $\tilde{V}_{ij}$ between carriers $i$ and $j$ are written as
\be
\tilde{V}_{ij}=V_{ij}+ \sum_{\ell} V_{i\ell}\tilde{\Pi}_{\ell\ell}\tilde{V}_{\ell j};\mbox{  } \tilde{V}_{ij}=\frac{V_{ij}}{\tilde\varepsilon},
\label{V dyson}
\ee
where $V_{ij}$ is  $v_q (\equiv \frac{4\pi e^2}{q^2})$ for $i=j$ or $-v_q$ for $i\neq j$.
The coupled equations of Eq. (\ref{V dyson}) for $\tilde{V}_{ij}$ also define the dielectric function $\varepsilon$ of the many carrier system written as
\be
\varepsilon (\vec q,\omega) = 1-v_q \tilde{\Pi}(\vec q,\omega).
\label{epsilon}
\ee
The equation of motion for $\tilde{\Pi}$ is not of closed form, and in the mean field or random phase approximation (RPA) \cite{Bruus,Mahan}, $\tilde{\Pi}$ is approximated by $\Pi^0$, the irreducible (Lindhard) polarization function of a noninteracting many carrier system \cite{Pi0}.
The latter $\Pi^{0}(\vec q,\omega)$ can be written in terms of its zero-temperature Lindhard function \cite{Maldague,Giuliani} to be employed in numerical analysis of the renormalized phonon propagators.
Hence, although $\Pi(\vec q,\omega,T)$ in Eq.(\ref{Pi dyson}) is a complex function, one can write the real and imaginary parts of $\Pi(\vec q,\omega)$ at finite temperature, in terms of $\mathcal{R}e \Pi^0$, $\mathcal{I}m \Pi^0$, and $\varepsilon(\vec q,\omega)$ \cite{Giuliani}.
In Figs. \ref{Pi_37K} and \ref{epsilon_37K}, the RPA $\Pi(\vec q,\omega)$ and $\varepsilon(\vec q,\omega)$ at 37 K are illustrated, respectively, for carrier concentrations of $10^{17} \rm cm^{-3}$ and $2 \times 10^{19} \rm cm^{-3}$. 
Pairs of dashed lines denote the region of allowed quasiparticle excitations in a wurtzite GaN. 
The zero of ${R}e \Pi (\vec q,\omega)$ is indicated by thin solid lines in Figs. \ref{Pi_37K}(a) and (c).
The zero of ${R}e ~\varepsilon (\vec q,\omega)$ are indicated, respectively, by thin solid lines in Fig. \ref{epsilon_37K}.

The phonon spectral function $\mathcal{A}(\vec q,\omega)$ is, now, given by
\begin{align}
\mathcal{A}& (\vec q,\omega)  \nonumber \\
= &\frac{4\omega_{\rm LO}^2 \Gamma(\vec q,\omega)/\pi}{[\omega^2-  \omega_{\rm LO}^2 -2 \omega_{\rm LO}\Delta(\vec q,\omega)]^2 + [2\omega_{\rm LO}\Gamma(\vec q,\omega)]^2 },
\label{Aqw}
\end{align}
where $\Delta$ and $\Gamma$, are written, respectively, as 
$\Delta(\vec q,\omega)= \mid M_{\vec q}\mid^2\mathcal{R}e\Pi(\vec q,\omega)/\hbar$ and $\Gamma(\vec q,\omega)=-\mid M_{\vec q}\mid^2\mathcal{I}m\Pi(\vec q,\omega)/\hbar$. 
The $\Delta$ describes the frequency renormalization correction due to the electronic screening of the long-ranged Coulomb fields associated with the phonons, and the $\Gamma$ is a measure of phonon lifetime $\tau_{\vec q}$ or the width of the spectral function due to collisional broadening \cite{Han2000}. (See the discussion below.)   
We note that $\mathcal{A}(\vec q,\omega)$ is peaked at $\omega=\omega_{\rm LO}\sqrt{1+\frac{ 2 \Delta}{\omega_{\rm LO}}}$. 
If one ignores the phonon frequency renormalization, $D$ reduces, with an infinitesimal positive $\eta$, to 
$D_0(\vec q,\omega)= {2\omega_{\rm LO}}/{[(\omega+i\eta)^2-  \omega_{\rm LO}^2]}$,
\label{bare phonon propagator}
resulting the bare phonon spectral function \cite{Mahan} 
$\mathcal{A}_0(\vec q,\omega)= [\delta(\omega-\omega_{\rm LO})- \delta(\omega+\omega_{\rm LO})]$.
\begin{figure*}[th]
\includegraphics[width=14cm]{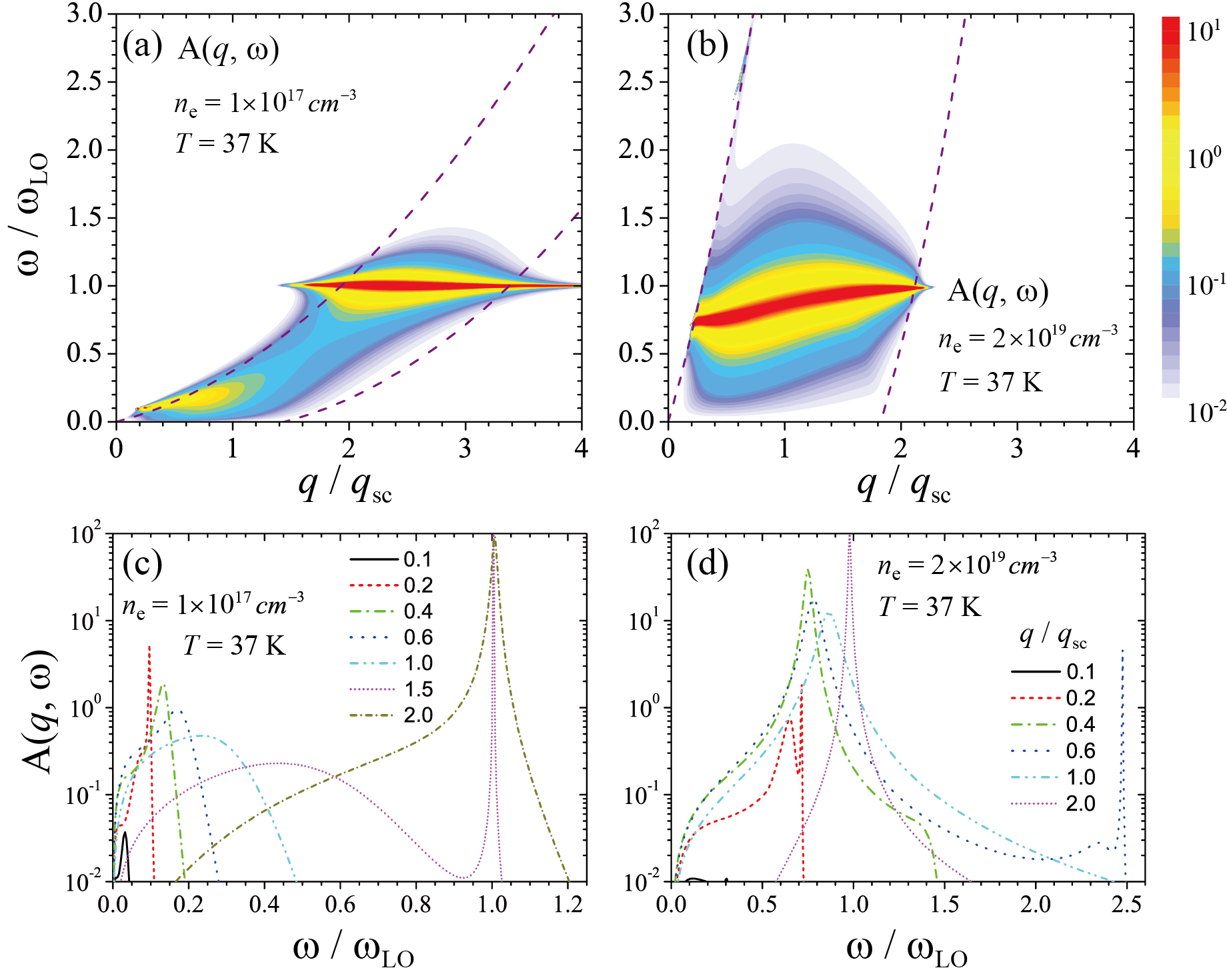}
\caption{Phonon spectral function $A(q,\omega)$ of a single component plasma formed by conduction electrons at carrier temperature 37 K.   
(a) carrier concentration of $10^{17} \rm cm^{-3}$. 
(b) carrier concentration of $2 \times 10^{19} \rm cm^{-3}$. 
(c) and (d) cross-sectional view of $A(q,\omega)$ at constant values of wavenumber $q$. 
}
\label{A}
\end{figure*}
Let us rewrite the denominator of ${D}(\vec q,\omega)$, in terms of phenomenological renormalized phonon frequency $\tilde{\omega}_{\vec q}$ and phonon lifetime $\tau_{\vec q}$, as 
$\omega^2- (\tilde{\omega}_{\vec q} -\frac{i}{\tau_{\vec q}})^2 $.
Then, $\tilde{\omega}_{\vec q}$ satisfies a quadratic equation given by
\be
\tilde{\omega}_{\vec q}^4 - \omega_{\rm LO}^2 (1+{ 2 \Delta}/{\omega_{\rm LO}})\tilde{\omega}_{\vec q}^2 -\omega_{\rm LO}^2\Gamma^2 =0
\label{ren omega_ph}
\ee
with
$
1/\tau_{\vec q}=\frac{\omega_{\rm LO}}{\tilde{\omega}_{\vec q}} \Gamma(\vec q,\omega)$. 
Here we note that the electron-phonon interaction introduces the phonon self-energy to change the bare phonon frequencies $\omega_{\rm LO}$ to the new frequencies $\tilde{\omega}_{\vec q}$ with finite lifetime $\tau_{\vec q}$.
Now the denominator of Eq. (\ref{Aqw}) is rewitten, in terms of $\tilde{\omega}_{\vec q}$, as
\be
{[\omega^2-  {\tilde{\omega}_{\vec q}}^2 + (\omega_{\rm LO}/\tilde{\omega}_{\vec q})^2\Gamma(\vec q,\omega)^2]^2 + [2\omega_{\rm LO}\Gamma(\vec q,\omega)]^2 },
\label{Aqw-1}
\ee
which determines the behavior of the spectral function $\mathcal{A}(\vec q,\omega)$ in the $\omega-q$ plane.

Solving Eq. (\ref{ren omega_ph}) for $\tilde{\omega}_{\vec q}$ gives rise to, in addition to the renormalized primary mode close to  
$\tilde{\omega}_{\vec q 1}\simeq \omega_{\rm LO} (1+\frac{ 2 \Delta}{\omega_{\rm LO}}+\frac{\Gamma^2/\omega_{\rm LO}^2}{1+ 2 \Delta/\omega_{\rm LO}})^{1/2}[\equiv{\Omega}_{1}(\vec q,\omega)]$ 
for $ 2 \Delta(\vec q,\omega) > -{\omega_{\rm LO}}$, a secondary mode of $\tilde{\omega}_{\vec q 2}\simeq {\Gamma}/{{\sqrt{|1+\frac{ 2 \Delta}{\omega_{\rm LO}}|}}}[\equiv{\Omega}_{2}(\vec q,\omega)]$ for $2\Delta(\vec q,\omega) < -\omega_{\rm LO}$ with negative self-energy correction $\Delta(\vec q,\omega) (<0)$.
The latter coupled mode would be well-defined only with finite values of collisional broadening $\Gamma(\vec q,\omega)$ ({\it i.e.}, $\mathcal{I}m\Pi(\vec q,\omega)<0$) over the region of $\mathcal{R}e\Pi(\vec q,\omega)<0$ in the $\omega-q$ plane. 
For the case $ 2 \Delta(\vec q,\omega) = -{\omega_{\rm LO}}$, Eq.(\ref{ren omega_ph}) is satisfied with  $\tilde{\omega}_{\vec q}=\sqrt{\omega_{\rm LO}\Gamma(\vec q,\omega)}$, leading us to
\[
\mathcal{A}(\vec q,\omega)=
\frac{4\omega_{\rm LO}^2 \Gamma(\vec q,\omega)/\pi}{\omega^4 + 4\omega_{\rm LO}^2\Gamma(\vec q,\omega)^2}.
\]
Deep valley of negative values of $\mathcal{R}e\Pi(\vec q,\omega)$ occurs just outside the quasiparticle excitation continuum in the $\omega-q$ plane and below the zero-value contour indicated by a thin solid line in Fig. \ref{Pi_37K}(a) and (c)].
We note that the secondary mode ${\Omega}_{2}$ is not expected in an approximation of taking $\mathcal{I}m\Pi(\vec q,\omega)=0$ and, hence, $\Gamma=0$ in the approximation \cite{DasSarma1988}.

\section{Results and Discussion}
In the calculation for numerical results, we use $m*= 0.22 m_0$ for the electron effective mass assuming a simple parabolic band of an ideal GaN material.
The bare plasma frequencies for carrier concentrations of $10^{17} \rm cm^{-3}$ and $2 \times 10^{19} \rm cm^{-3}$ are $\omega_{\rm p} \sim 10.8 \rm meV ~(= 0.12\omega_{\rm LO})$ and $\omega_{\rm p} \sim 150 \rm meV ~(= 1.7\omega_{\rm LO})$, respectively, for the plasmas of conduction electrons.  

The spectral behaviors of  $\Pi(q,\omega)$ and $\varepsilon(q,\omega)$ are modified from that of noninteracting counter parts, $\Pi^0(q,\omega)$ and $\varepsilon_0(q,\omega)$, as illustrated in Figs. \ref{Pi_37K} and \ref{epsilon_37K}, revealing the character of the optic plasmonic branches outside the region of quasiparticle excitation continuum \cite{pi0-2}.
In both figures, the wavenumber and frequency are displayed in units of the Thomas-Fermi screening wavenumber $q_{\rm sc}$ and the bare LO phonon frequency $\omega_{\rm LO}$.
The $q_{sc}$ is a decreasing function of temperature and $q_{sc} \simeq 1.39 k_F$ ($1.01 k_F$) for carrier concentration of $n =10^{17} \rm cm^{-3}$ ($n =2\times 10^{19} \rm cm^{-3}$) at 37 K. 
[The zero of $\mathcal{R}e \Pi_c^0 (q,\omega)$ lies in the continuum region of the single-particle excitations, and, at small $q$, $\mathcal{R}e \Pi_c^0 (q,\omega)$ changes sign from negative to positive as $\omega$ increases sweeping across the continuum region. 
On the other hand, $\mathcal{I}m \Pi_c^0 (q,\omega)$ is finite and negative in the continuum region of the single-particle excitations showing peak structure near the zero line of  $\mathcal{R}e \Pi_c^0 (q,\omega)$.
This observation is a direct consequence of the Kramers-Kronig dispersion relations \cite{Giuliani}. 
Within the continuum region, electrons within the  Fermi sea can be excited to states outside the Fermi sea. 
The behavior of $\Pi_c^0 (q,\omega)$ means that single-particle excitations of free carriers are the only processes for the energy and momentum dissipation, because the effects of carrier screening is completely ignored in $\Pi_c^0 (q,\omega)$.] 
Thin solid lines in Fig. \ref{epsilon_37K} show the contours of $\varepsilon(q,\omega)=0$, which denote the dispersion curves of the longitudinal electronic plasmons with plasma cut-offs at $(\omega_{\rm c}, q_{\rm c})\simeq (0.14\omega_{\rm LO},0.5q_{\rm sc})$ and $(2.6\omega_{\rm LO},0.7q_{\rm sc})$ for carrier concentrations of $10^{17} \rm cm^{-3}$ and $2 \times 10^{19} \rm cm^{-3}$, respectively. 
The well defined optic plasmon modes are intact to be clearly seen, occurring well outside the quasiparticle excitation continuum, in the plot of $\mathcal{R}e ~\varepsilon(q,\omega)$ at frequencies lower (higher) than that of the bare LO phonons for carrier concentrations of $10^{17} \rm cm^{-3}$ ($2 \times 10^{19} \rm cm^{-3}$).

In Fig. \ref{A}, the spectral behaviors of the phonon spectral functions $A(q,\omega)$ are illustrated for carrier concentrations of $1 \times 10^{17} \rm cm^{-3}$ and $2 \times 10^{19} \rm cm^{-3}$ at carrier temperature 37 K.
In panel (c) and (d) cross-sectional view of $A(q,\omega)$ are illustrated for several different values of $q$. 
The dispersion relations of the peaks of $A(q,\omega)$ in the $\omega-q$ plane represent the renormalized phonon--plasmon (coupled) mode branches with collisional broadening.
The mode coupling of the LO phonons and plasmons introduces a pair of branches named $L^{(+)}(\omega,q)$ and $L^{(-)}(\omega,q)$, the former (latter) representing high-frequency (low-frequency) mode \cite{Cochran,Cohen}.

For carrier concentration $n =10^{17} \rm cm^{-3}$ in n-doped GaN, the case shown in panel (a) and (c), $\omega_{\rm p} \ll \omega_{\rm LO}$, since $\omega_{\rm p} \sim 10.8 \rm meV (\simeq 0.12 \omega_{\rm LO})$, while, for carrier concentration $n =2 \times 10^{19} \rm cm^{-3}$, the case shown in panel (b) and (d), $\omega_{\rm p} \gg \omega_{\rm LO}$, since $\omega_{\rm p} \sim 150 \rm meV (\simeq 1.7 \omega_{\rm LO})$. 
%

For the case of highly doped plasma coupled with LO phonons [Fig. \ref{A}(b)], the $L^{(+)}(\omega,q)$ branch is highly plasmon-like with peaks of large dispersion dominant near $\omega \simeq 2.5 \omega_{\rm LO}$, while the $L^{(-)}(\omega,q)$ mode shows strong phonon-like behavior with broad peaks for frequencies lying between $\omega_{\rm TO} < \omega <\omega_{\rm LO}$. 
The $L^{(-)}(\omega,q)$ branch shows the behavior of highly screened long-ranged Coulombic phonon fields by the high frequency plasmonic carriers for $q \leq q_{\rm sc}$ and ineffectiveness of the carrier screening to the phonon fields resulting in the bare phonon frequency $\omega_{\rm LO}$ for $q \geq q_{\rm sc}$.
Within the quasiparticle excitation continuum region, modes are ill-defined because they are subject to Landau damping. 
Beyond the continuum at large wave numbers, the screening is ineffective and, hence, the effectively bare modes are resumed with sharp peaks.
The frequency of the coupled LO modes tends to $\omega_{\rm LO}$ for $q \gg q_{\rm sc}$.
On the other hand, for the case of lightly doped plasma coupled with LO phonons [Fig. \ref{A}(a)], the $L^{(+)}(\omega,q)$ branch is highly phonon-like with negligible screening effect by the low frequency plasmonic carriers, but the $L^{(-)}(\omega,q)$ mode shows plasmon-like behavior with broad peaks. 

The plasmon-phonon coupled modes $L^{(\pm)}(\omega,q)$ show relatively narrow peaks in the phonon spectral function $A(q,\omega)$, while the QPE-like modes shows strong dissipative behavior in $A(q,\omega)$ as seen in Fig. \ref{A}(c) and (d).
For weakly doped plasma with carrier concentration $10^{17} \rm cm^{-3}$ [Fig. \ref{A}(c)], the maximum in $A(q,\omega)$ of the phonon-like $\Omega_+(q)$ coupled branch shows less dispersive unscreened behavior with frequencies $\sim \omega_{\rm LO}$, because the low frequency plasma species is not fast enough to screen the long-ranged Coulombic part of the LO oscillation.
The low frequency plasmon-like $\Omega_-(q)$ coupled branch reveals slight dispersive behavior starting with $\omega \sim \omega_{\rm p} (\simeq 0.12\omega_{\rm LO})$, while the QPE-like modes shows heavy broadening as increasing the wavenumber.
Therefore, for small frequency exchanges during carrier-carrier scattering, the finite-$\omega$ (dynamic) effect of the screening is negligible reducing to the case of Thomas-Fermi screening limit.
  
For heavily doped plasma with carrier concentration $2 \times 10^{19} \rm cm^{-3}$ [Fig. \ref{A}(d)], the maximum in $A(q,\omega)$ of the phonon-like $\Omega_-(q)$ coupled branch appears, as increasing wavenumber, at frequencies between $\omega_{\rm TO}$ and $\omega_{\rm LO}$ subject to Landau damping approaching asymptotically to $\omega_{\rm LO}$ \cite{Abstreiter1984}.
The long-ranged Coulombic field associated with the longitudinal optical phonons is strongly screened by the carriers of high plasmonic frequency $\omega_{\rm p}~ (\simeq 1.7 \omega_{\rm LO})$.  
The higher frequency plasmon-like $\Omega_+(q)$ coupled branch starts with $\omega$ well above $\omega_{\rm p}$ showing strong dispersive behavior of relatively weaker strength.  %

\begin{figure*} 
\includegraphics[width=14cm]{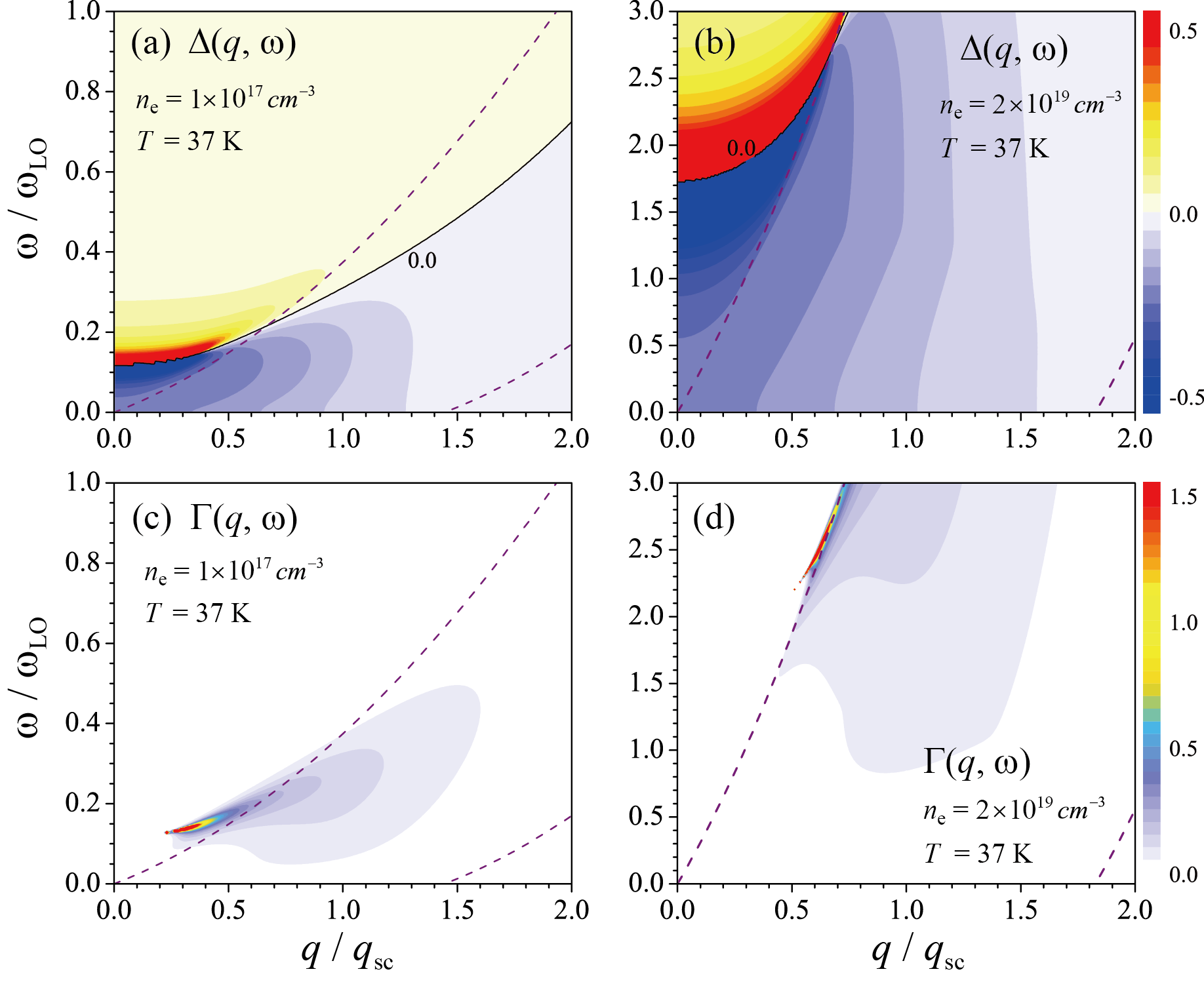}
\caption{Phonon self energy of a conduction electron plasma in a wurtzite GaN at 37 K.
(a) and (b) real parts of the self energy correction $\Delta(\vec q,\omega)$ for carrier concentrations of $10^{17} \rm cm^{-3}$ and $2 \times 10^{19} \rm cm^{-3}$.
(c) and (d) imaginary parts of the self energy correction $\Gamma(\vec q,\omega)$ for carrier concentrations of $10^{17} \rm cm^{-3}$ and $2 \times 10^{19} \rm cm^{-3}$.
}
\label{SE}
\end{figure*}
\begin{figure*} 
\includegraphics[width=14cm]{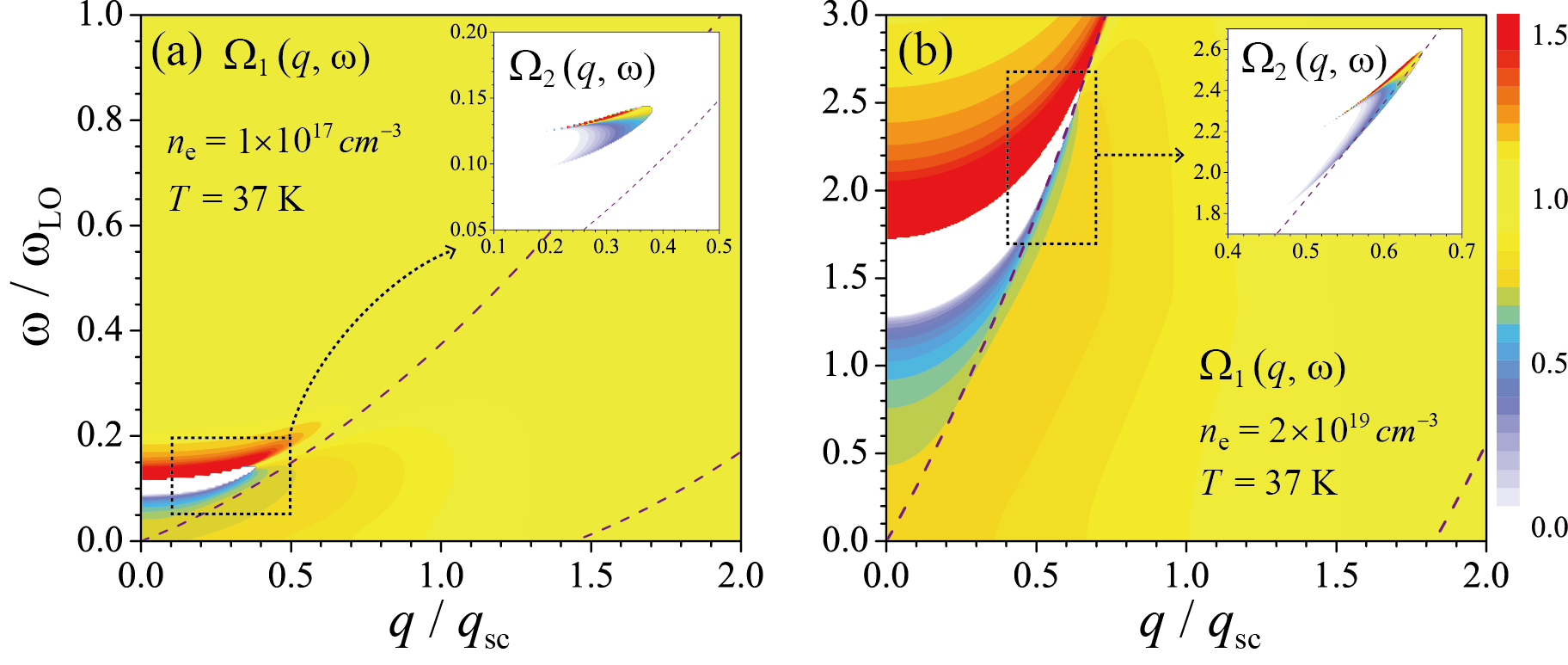}
\caption{Spectral behavior of the plasmon-phonon coupled modes in a doped wurtzite GaN at 37 K.
Frequency and wavenumber dependences of $\Omega_1(q,\omega)$ and $\Omega_2(q,\omega)$ for electron concentrations of (a) $10^{17} \rm cm^{-3}$ and (b) $2 \times 10^{19} \rm cm^{-3}$.} 
\label{omega}
\end{figure*}
\begin{figure*} 
\includegraphics[width=14cm]{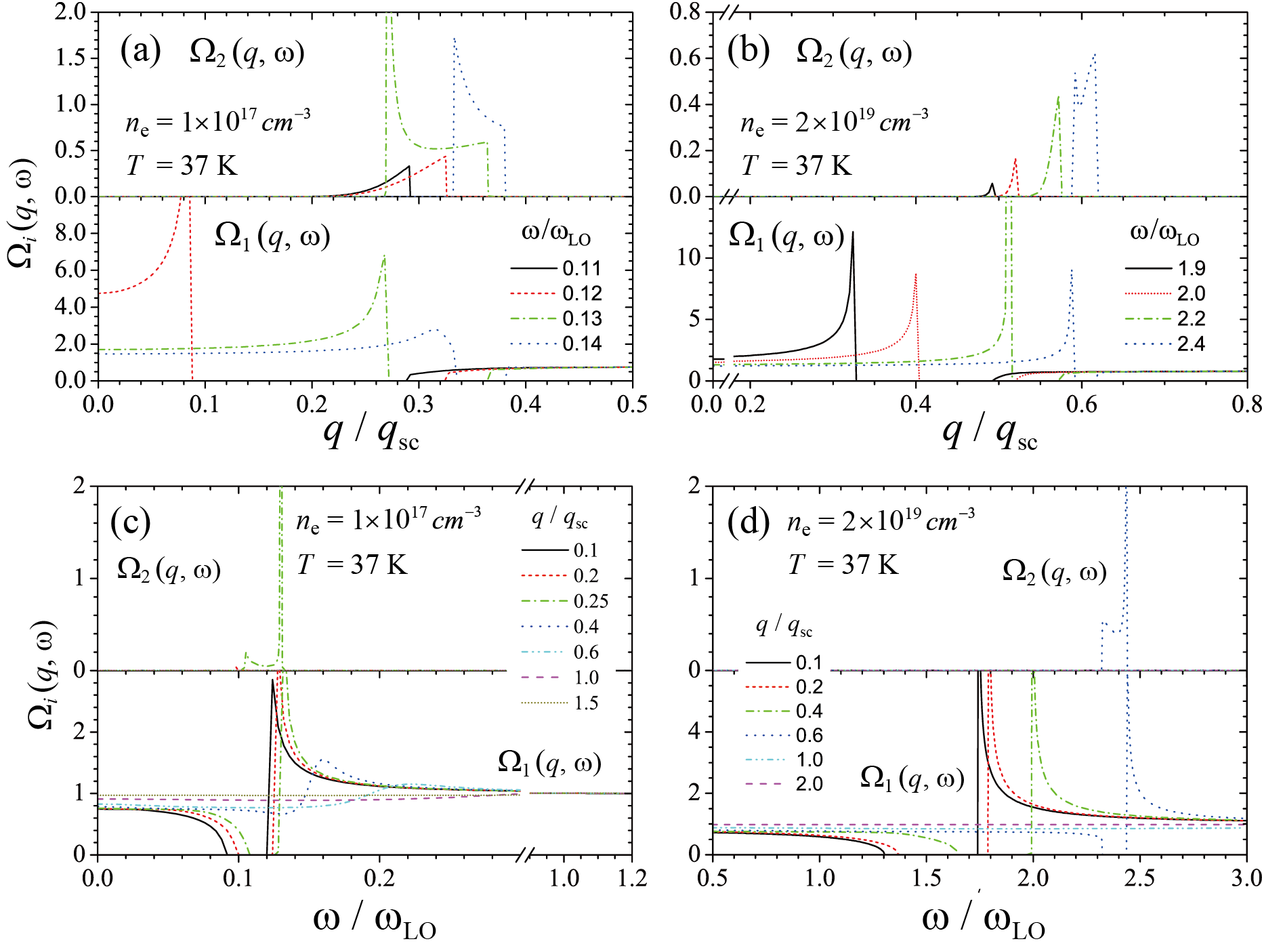}
\caption{Cross-sectional view of renormalized phonon modes $\Omega_1(q,\omega)$ and $\Omega_2(q,\omega)$ at 37 K
Behavior at constant values of frequency $\omega$ for electron concentrations of (a) $10^{17} \rm cm^{-3}$ and (b) $2 \times 10^{19} \rm cm^{-3}$.
Behavior at constant values of wavenumber $q$ for electron concentrations of (c) $10^{17} \rm cm^{-3}$ and (d) $2 \times 10^{19} \rm cm^{-3}$.
}
\label{SEq-SEw}
\end{figure*}

In Fig. \ref{SE}, the real part $\Delta(\vec q,\omega)$ and imaginary part $\Gamma(\vec q,\omega)$ of the phonon self-energy correction at 37 K are illustrated for carrier concentrations of $10^{17} \rm cm^{-3}$ and $2 \times 10^{19} \rm cm^{-3}$.
The signs of $\Delta(\vec q,\omega)$ the same as that of ${R}e\Pi(\vec q,\omega)$, and the contours of $\Delta(q,\omega)=0$ are indicated by thin solid lines in Fig. \ref{SE}(a) and (b). 
On the other hand, $\Gamma(\vec q,\omega)$ is defined to be nonnegative and shows peaked structure along the well-defined plasmon dispersion curves near the plasma cut-offs as seen in Fig. \ref{epsilon_37K}. 
We note that LO phonon damping is dominated through emission of longitudinal plasmons along with weak contribution from QPEs of relatively broader distribution.   

In Fig. \ref{omega}, the frequency and wavenumber dependences of the plasmon-LO phonon coupled modes $\Omega_1(q,\omega)$ and $\Omega_2(q,\omega)$ are shown for electron concentration of $10^{17} \rm cm^{-3}$ and $2 \times 10^{19} \rm cm^{-3}$. 
We note that $\Omega_1(q,\omega)\simeq \omega_{\rm LO}$ over most of the domain satisfying the condition  $2 \Delta(\vec q,\omega) > -{\omega_{\rm LO}}$ in the $\omega-q$ plane, except near the regions of well-defined plasmonic collective modes becoming $\Omega_1(q,\omega) >> \omega_{\rm LO}$.
Below the small opening gap [as indicated with white blank in panel (a)] of $2 \Delta(\vec q,\omega) < -{\omega_{\rm LO}}$, we observe low frequency of $\Omega_1(q,\omega) < \omega_{\rm LO}$ over the region designated in green. 
The secondary mode $\Omega_2(q,\omega)$ is confined in the region of $2 \Delta(\vec q,\omega) < -{\omega_{\rm LO}}$ with finite values of $\Gamma(\vec q,\omega)$, as shown in Fig. \ref{SE}(c) and (d). 
The coupled modes reveal strong frequency and wavenumber dependences, and cross-sectional views of the renormalized phonon frequencies $\Omega_1(q,\omega)$ and $\Omega_2(q,\omega)$ are illustrated in Fig. \ref{SEq-SEw} for constant values of frequency $\omega$ in panels (a) and (b) and for constant values of wavenumber $q$ in panels (c) and (d), respectively, at 37 K for electron concentration of $10^{17} \rm cm^{-3}$ and $2 \times 10^{19} \rm cm^{-3}$.


\vspace{-0.5cm}
\section{Summary and Conclusion}
In this paper, spectral analysis on the phonon spectral functions of a solid-state plasma formed in a doped semiconductor are investigated by examining phonon self-energy corrections within a random phase approximation.
Phonon spectral functions are mapped out in the $\omega-q$ space, and their dynamic and nonlocal behaviors are studied including the effects of dynamic screening and plasmon-phonon coupling at finite temperature. 
The results are applied to the case of a wurtzite GaN considering carrier-phonon channel of polar optical phonons.
We presented the frequency and wavenumber dependent behavior of well-defined plasmon-dominated modes clearly seen well outside the single particle excitation continuum and the optical phonon-dominated branch.

Our result shows that, in the presence of dynamic screening of many carriers, the phonon spectral function is drastically modified from the bare phonon case through electron-LO phonon coupling channel. 
We also find that the contribution of the QPE-like mode to the spectral function is heavily dependent on the wavenumber and the concentration of plasma species in a doped semiconductor.
For the case of heavily doped electron plasma, we confirm that the maximum in $A(q,\omega)$ of the phonon-like $\Omega_1(q)$ coupled branch appears, as increasing wavenumber, at frequencies between $\omega_{\rm TO}$ and $\omega_{\rm LO}$ subject to Landau damping approaching asymptotically to $\omega_{\rm LO}$.
The long-ranged Coulombic field associated with the longitudinal optical oscillations is completely screened by the plasma species of $\omega_{\rm p}~ (\gg \omega_{\rm LO})$ with carrier concentration $2 \times 10^{19} \rm cm^{-3}$. 
The higher frequency plasmon-like coupled branch $\Omega_+(q)$ starts at $q=0$ with $\omega \geq \omega_{\rm p} (\simeq 1.7\omega_{\rm LO})$ showing strong dispersive behavior of relatively weaker strength.  
For the case of weakly doped electron plasma, the maximum in $A(q,\omega)$ of the high frequency phonon-like $\Omega_1(q)$ coupled branch shows non-dispersive (unscreened) behavior with frequencies $\sim \omega_{\rm LO}$.
The plasma species is not fast enough to screen the long-ranged Coulombic part of the LO oscillation
The low frequency plasmon-like branch $\Omega_-(q)$ starts with $\omega_{\rm p} (\sim 0.12\omega_{\rm LO})$. 
For small frequency exchanges during carrier-carrier scattering, the nonlocal effect of the screening is reduced to the case of Thomas-Fermi screening limit.

Meaningful informations on the collective behavior of plasmon-phonon coupled system with consideration of thermal and collisional broadening effects can be obtained by comparing experimental spectra on polar semiconductors with the results presented in the present paper.
The screening effects on the phonon spectral function presented in the present work would be resolved in various experiments such as transport measurements and scattering experiments with neutrons for finite wavenumber exchanges and Raman techniques in the region of vanishing wavenumber exchanges.

\begin{acknowledgments}
One of the authors (KSY) acknowledges a support in part by a two-year grant from the PNU Research Foundation.
\end{acknowledgments}


\begin{thebibliography}{99}


\bibitem{Abstreiter1984} G. Abstreiter, M. Cardona, and A. Pinczuk, {\it Light Scattering in Solids IV} ed. by M. Cardona and G. G\"untherodt, (Springer-Verlag, New York, 1984), p.5, and the references there in.
\bibitem{Pinczuk1977} For example, A. Pinczuk, G. Abstreiter, R. Trommer, and M. Cardona, Solid St. Commun.  {\bf 21}, 959 (1977), and the references therein.
\bibitem{Abstreiter1979} G. Abstreiter, R. Trommer, M. Cardona, and A. Pinczuk, Solid St. Commun.  {\bf 30}, 703 (1979), and the references therein.
\bibitem{Romanek} K.M. Romanek, H. Nather, and E.O. G\"obel,  Solid St. Commun. {\bf 39}, 23 (1981).
\bibitem{Quinn-Yi} J.J. Quinn and K.S. Yi, {\it Solid State Physics}, (Springer, Heidelberg, 2009), Ch. 8.

\bibitem{Jain1988} J.K. Jain, R. Jalabert, and S. Das Sarma, Phys. Rev. Lett. {\bf 60}, 353 (1988).


\bibitem{Bruus}  H. Bruus and K. Fensberg, {\it Many-Body Quantum Theory in Condensed Matter Physics}, (Oxford Univ. Press, New York, 2004).
\bibitem{Mahan} See, for example, G. Mahan, {\it Many Particle Physics}, 3rd ed. (Plenum, 
New York, 2000).
\bibitem{Mahan72} G. Mahan, {\it Polarons in Ionic Crystals and Polar 
Semiconductors}, edited by J.T. Devreese (North Holland, Amsterdam, 1972), pp. 553.
\bibitem{ksyi2007} K.S. Yi, J.S. Kim, and K. Kyhm, J. Korean Phys. Soc. {\bf 50}, 1670 (2007).
\bibitem{Mq} Explicit expression of the matrix element for LO phonons is given by $
|M_{q}(q)|^2 = \frac{4\pi \hbar}{q^2} \sqrt{\frac{\hbar^3\omega_{\rm LO}^3}{2m_\nu}} \alpha,
$
where $\alpha$ is the dimensionless Fr\"ohlich coupling constant $\alpha=\frac{e^2}{\hbar}\sqrt{m_{\nu}/(\hbar\omega_{\rm LO})}\left(\frac{1}{\varepsilon_\infty}-\frac{1}{\varepsilon_0}\right)$.

\bibitem{Pi0} The Lindhard polarization function is written, with infinitesimally positive $\eta$, as
$\Pi^{0}(\vec q,\omega) = -\frac{2}{\hbar}\sum_{\vec k}\frac{f_{k+q}^{(0)}-f_{k}^{(0)}}{\omega-(\omega_{k+q}^{(0)}-\omega_{k}^{(0)})+i\eta}$, where $f_{k}^{(0)}$ and $\omega_{k}^{(0)}$ are the finite temperature carrier distribution function and carrier energy of  wavenumber $k$ in the plasma. 

\bibitem{Giuliani} G. Giuliani and G. Vignale, {\it Quantum Theory of Electron Liquid}, (Cambridge Univ. Press, New York, 2005).
\bibitem{Maldague} P.F. Maldague, Surf. Science {\bf 73}, 296 (1978).

\bibitem{Han2000} See, for example, J.E. Han and O. Gunnarsson, Phys. Rev. B {\bf 61}, 8628 (2000).

\bibitem{DasSarma1988} S. Das Sarma, J.K. Jain, and R. Jalabert, Phys. Rev. B {\bf 37}, 6290 (1988).

\bibitem{pi0-2}  The noninteracting polarization function $\mathcal{R}e \Pi^0 (q,\omega)$ shows a broad dip structure within the single particle excitation continuum for $q < q_{\rm sc}$ at small frequency $\omega \ll \omega_{\rm LO}$ and a broad peak structure along the upper boundary of the continuum, while $\mathcal{I}m \Pi^0 (q,\omega)$ shows a sharp dip structure inside the single particle excitation continuum. 

\bibitem{Cochran} See, for example, W. Cochran, R.A. Cowley, G. Dolling, and M.M. Elcombe, Proc. Roy. Soc., Ser. A {\bf 293}, 433 (1966). The authors illustrated the dispersion relation of the coupled modes.
\bibitem{Cohen} See, for example, M. Cohen, in {\it Superconductivity}, edited by R.D. Parks (Dekker, New York, 1969), p.659. The author discussed the behavior of the coupled modes neglecting the damping effect in the long wavelength limit.

\end{thebibliography}
\end{document}